\documentclass[%
 reprint,
 amsmath,amssymb,
 aps,
]{revtex4-2}

\usepackage{graphicx}
\usepackage{dcolumn}
\usepackage{bm}

\begin{document}

\preprint{APS/123-QED}

\title{Megastable quantization in generalized pilot-wave hydrodynamics}

\author{Álvaro G. López$^1$}\email{alvaro.lopez@urjc.es}
\author{Rahil N. Valani$^2$}%
 
\affiliation{$^1$Nonlinear Dynamics, Chaos and Complex Systems Group.\\Departamento de F\'isica, Universidad Rey Juan Carlos, Tulip\'an s/n, M\'ostoles, 28933, Madrid, Spain}

\affiliation{$^2$Rudolf Peierls Centre for Theoretical Physics, Parks Road,
University of Oxford, OX1 3PU, United Kingdom}

\date{\today}

\begin{abstract}

A classical particle in a harmonic potential gives rise to a continuous energy spectra, whereas the corresponding quantum particle exhibits countably infinite quantized energy levels. In recent years, classical non-Markovian wave-particle entities that materialize as walking droplets have been shown to exhibit various hydrodynamic quantum analogs, including quantization in a harmonic potential by displaying few coexisting limit cycle orbits. By considering a truncated-memory stroboscopic pilot-wave model of the system in the low dissipation regime, we obtain a classical harmonic oscillator perturbed by oscillatory non-conservative forces that displays countably infinite coexisting limit-cycle states, also known as \emph{megastability}. Using averaging techniques in the low-memory regime, we derive analytical approximations of the orbital radii, orbital frequency and Lyapunov energy function of this megastable spectrum, and further show average energy conservation along these quantized states. Our formalism extends to a general class of self-excited oscillators and can be used to construct megastable spectrum with different energy-frequency relations. 

\end{abstract}

\maketitle


\textit{Introduction.} Orbit quantization is largely considered an exclusive feature of microscopic quantum systems. However, recent studies with a classical hydrodynamic system of millimeter-sized walking oil droplets~\citep{Couder2005} have demonstrated several hydrodynamic quantum analogs~\citep{Bush2020review,Bush2024,10.1063/5.0210055}, including an analog of quantization~\citep{Perrard2014a}. In experiments, the droplet walks, while bouncing periodically, on a vibrating bath of the same liquid. Each bounce of the droplet imprints a localized standing wave on the liquid surface that decays slowly in time. The droplet in turn interacts with these self-generated waves on subsequent bounces resulting in horizontal self-propulsion. Three key features that make this system dynamically rich: (i) the droplet and its wave co-exist as a \emph{wave-particle entity} (WPE); without the droplet, surface waves decay completely, (ii) the system is \emph{non-Markovian} since the droplet retains \emph{path memory} from the slowly decaying self-generated waves that sculpt its complex dynamical landscape, and (iii) the system is \emph{active} in the sense of active matter~\citep{PismenActiveMatterBook} because the droplet locally extracts energy from the vibrating bath and converts it into horizontal motion.

The hydrodynamic analog of quantization is experimentally realized by confining the WPE in a two-dimensional harmonic potential, which results in a discrete but finite set of periodic orbits such as circles, ovals, lemniscates and trefoils, as the width of the harmonic potential is varied~\citep{Perrard2014a,Kurianskiharmonic,labousse2016b,Blits24}. The time-delayed feedback produced by the slowly decaying particle-generated waves constrain the particle's trajectory, resulting in quantum-like eigenmodes.
In both experimental and theoretical studies of a WPE in a harmonic potential (or in general confining potentials), one observes at most a few coexisting quantized periodic orbits~\citep{Perrard2014a,Perrard2014b,Tambasco2016,Tambascoorbit,Labousse_2014,labousseharmonic,Kurianskiharmonic,durey2018,Perrard2018}, whereas the corresponding quantum particle in a harmonic potential has an infinite spectrum of eigenstates.


In dynamical systems literature, the notion of a countably infinite structure of stable periodic orbits exists and has been dubbed \emph{megastability}. It was first numerically reported for periodically driven nonlinear oscillators~\cite{Sprott2017}. In this letter, we investigate the megastable structure of stable limit cycles emerging in the very low dissipation regime of a truncated-memory pilot-wave model of a walking droplet in a harmonic potential. 


 \begin{figure}
\centering
\includegraphics[width=0.7\columnwidth]{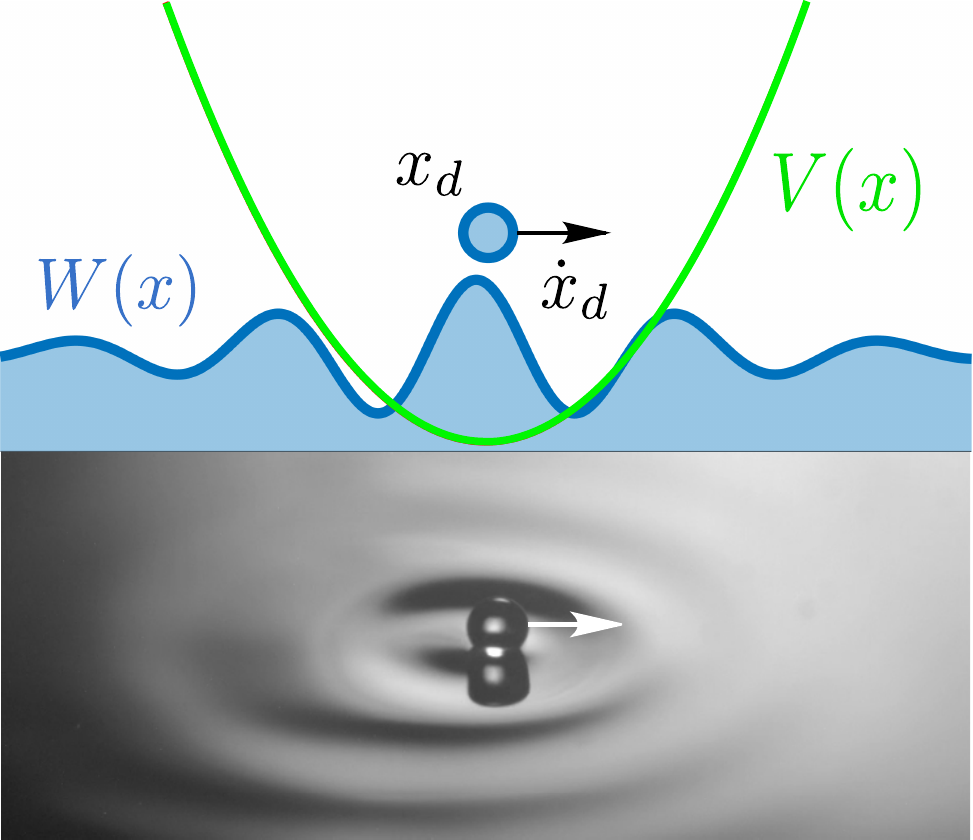}
\caption{(Top) Schematic of the hydrodynamic pilot-wave system. A particle located at position $x_d$ at time $t$ moves horizontally in one-dimension with velocity $ \dot{x}_d$ while continuously generating waves of spatial form $W(x)$ that decay exponentially in time. The particle experiences a force from external harmonic potential $V(x)$, a dissipative drag force and a propulsive force from its past self-generated waves up to time $t-t_c$. (Bottom) Experimental image of the hydrodynamic pilot-wave system showing a superwalking droplet~\citep{superwalker}.}
\label{Fig: schematic}
\end{figure}

\textit{Truncated-memory pilot-wave model.} By averaging over the fast vertical periodic bouncing motion of the walking droplet, \citet{Oza2013} developed a stroboscopic model that represents the walking droplet as a continuous emitter of damped standing waves and results in an integro-differential trajectory equation for the horizontal walking motion. We use a one-dimensional setup as illustrated schematically in Fig.~\ref{Fig: schematic}.  Consider a drop (particle) of mass $m$ located at $x_d$ at time $t$ and moving horizontally in one-dimension with velocity $\dot{x}_d$ in the presence of an external harmonic potential $V(x)=kx^2/2$ with spring constant $k$. At each instance of time, the particle generates a standing wave with spatial form $ W(x)$ centered at the particle position and the wave decays exponentially in time. As opposed to the standard stroboscopic model that takes into account the effects of wave memory up to the present time $t$~\citep{Oza2013}, here we consider a modified stroboscopic model where the particle experiences a wave-memory force from its self-generated waves up to time $t-t_c$ with $t_c$ as the cut-off memory time. The integro-differential trajectory equation of this one-dimensional WPE is given by~\citep{phdthesismolacek,Oza2013,ValaniUnsteady}: 
\begin{align}\label{Eq: dimensional eq}
    &m\ddot{x}_d+D\dot{x}_d + k x_d = \\ \nonumber
    &-\frac{F}{T_F}\int_{-\infty}^{t-t_c} W'\left(k_F(x_d(t)-x_d(s))\right)\,\text{e}^{-\frac{(t-s)}{T_F \text{Me}}}\,\text{d}s. 
\end{align}
On the left side of Eq.~\eqref{Eq: dimensional eq}, there are three terms - the first representing the inertia of the particle, the second representing the effective drag force on the particle and the third term is the force from the harmonic potential; where dots denote derivatives with respect to time. The term on the right side of the equation represents the force exerted on the particle by its self-generated wave field. The wave field is a superposition of the individual exponentially time-decaying waves of spatial form $W(x)$ that are generated by the particle continuously along its trajectory. This force is proportional to the gradient of the underlying wave field. The model in Eq.~\eqref{Eq: dimensional eq} reduces to the standard stroboscopic model~\citep{Oza2013} for $t_c=0$. Other physical parameters in Eq.~\eqref{Eq: dimensional eq} are as follows: $D$ is an effective drag coefficient, $k_F$ is the Faraday wavenumber,
$F=m g A k_F$ is a non-negative wave-memory force coefficient where $g$ is the gravitational acceleration and $A$ is the amplitude of droplet-generated surface waves, $\text{Me}$ is the memory parameter that describes the proximity to the Faraday instability, and $T_F$ is the Faraday period of droplet-generated standing waves and also the bouncing period of the walking droplet~(see \citet{Oza2013} for more details).

\textit{Low-memory self-excited oscillator model.} Non-dimensionalizing Eq.~\eqref{Eq: dimensional eq} by rescaling time with $T_F$ and space with $k_F$, and considering the low-memory regime where the particle is only influenced by its most recent wave at time $t-t_c$, we get~(see Appendix~A for derivation)
\begin{align}\label{Eq: dimless eq}
\ddot{x}_d+\mu\dot{x}_d+K x_d=-\epsilon W'(\tau\dot{x}_d).
\end{align}
Here the dimensionless parameters are given by: $\mu=DT_F/m$ as the drag coefficient, $K=k T_F^2/m$ as the spring constant, $\tau=t_c/T_F$ as the cut-off memory time and $\epsilon=\text{e}^{-\tau/\text{Me}} F k_F T_F^2/m$, as the wave force coefficient. For analytical tractability and clarity of presentation, we consider this low-memory model with a simple cosine wave form $W(x)=\cos(x)$ \citep{phdthesismolacek,Durey2020lorenz,ValaniUnsteady,Valanilorenz2022} for the remainder of the letter,
however, we highlight that the existence of megastability is robust in stroboscopic models with truncated memory and extends to larger wave-memory~(see Appendixes~B and C) and also alternate wave forms $W(x)$ with both oscillations and spatial decay~(see Appendixes~D and E). For the remainder of the letter, we fix $\tau=1$ which corresponds to cut-off memory time duration equivalent to that of a single bounce of the droplet. 

In phase space, we rewrite our dynamical system in Eq.~\eqref{Eq: dimless eq} as
\begin{align} \label{eq:5}
\dot{x}_d & = v_d, \\ \nonumber
\dot{v}_d & = - \mu v_d +\epsilon \sin v_d - x_d,
\end{align}
where we have rescaled $(x_d,t,\epsilon,\mu) \rightarrow (x_d/\sqrt{K}, t/\sqrt{K} , \epsilon \sqrt{K}, \mu \sqrt{K})$ to simplify the equations further. 

\textit{Megastable spectrum.} The system in Eq.~\eqref{eq:5} displays an unbound set of nested limit cycles and we use the Krylov-Bogoliubov averaging method to obtain analytical expressions for this spectrum. Physically, the megastable quantization results from a dynamical balance of forces on the particle that arise from the external harmonic potential and the self-generated waves.
The Krylov-Bogoliubov method~\citep{krylov1950} proposes the phase space ansatz $(x_d(t),v_d(t))=(r(t) \sin(t+\varphi(t)),r(t) \cos(t+\varphi(t)))$, and averages over the phase variable $\theta(t) = t+\varphi(t)$, using the fact that for small $\mu$ and $\epsilon$, there is a time-scale separation between the variations of the phase $\theta(t)$ and the amplitude $r(t)$ of the oscillation. The error of this approximation is bounded and can be made arbitrarily small by decreasing the size of $\max(\varepsilon,\mu)$~\citep{krylov1950}. Substituting the ansatz in Eq.~\eqref{eq:5}, we obtain the following system of differential equations for the amplitude and the phase of the orbit
\begin{align}
\dot{r} & = - (\mu r \cos \theta -\epsilon \sin (r \cos \theta)) \cos \theta , \label{eq:61} \\ 
r \dot{\varphi} & = (\mu r \cos \theta-\epsilon \sin (r \cos \theta)) \sin \theta. 
\label{eq:62}
\end{align}
Averaging Eqs.~\eqref{eq:61} and \eqref{eq:62} over the period $\theta \in [0,2\pi]$ of the unperturbed ($\mu=\epsilon=0$) harmonic oscillator yields $\dot{\varphi}=0$ and
\begin{equation}
\dot{r}  = -  \bar{\mu} r+\epsilon J_{1} (r),
\label{eq:7}
\end{equation}
where $J_{1}(r)$ is the first order Bessel function of the first kind, and we have defined the parameter $\bar{\mu} = \mu/2$, to simplify the equation further. In the particular case of no dissipation, $\bar{\mu}=0$, the fixed points of Eq.~\eqref{eq:7}, and hence the amplitudes of the limit cycles, are given by the infinite roots of the Bessel function which for large $r$ have the asymptotic relation $J_{1}(r) \approx \sqrt{\pi/2r} \cos(r-3\pi/4)$. Thus, we have the existence of a countably infinite sequence of limit cycles with alternating stability, with the amplitude sequence converging asymptotically to the zeros of the cosine function, $r_n \approx \pi(n+5/4)$, where $r_n$ denotes the $n$th root of Eq.~\eqref{eq:7}. We note that the dynamical system in Eq.~\eqref{eq:5} with $\bar{\mu}=0$ has been previously explored as a mathematical curiosity~\citep{Kahn2014-na}, and attempts have been made to make vague connections with quantum eigenstates~\citep{ZARMI201721}. However, these were not rooted in a physical system, whereas in our pilot-wave walking-droplet system, these equations emerge in the low-memory limit of the droplet's integro-differential equation of motion.
 \begin{figure}
\centering
\includegraphics[width=0.9\columnwidth]{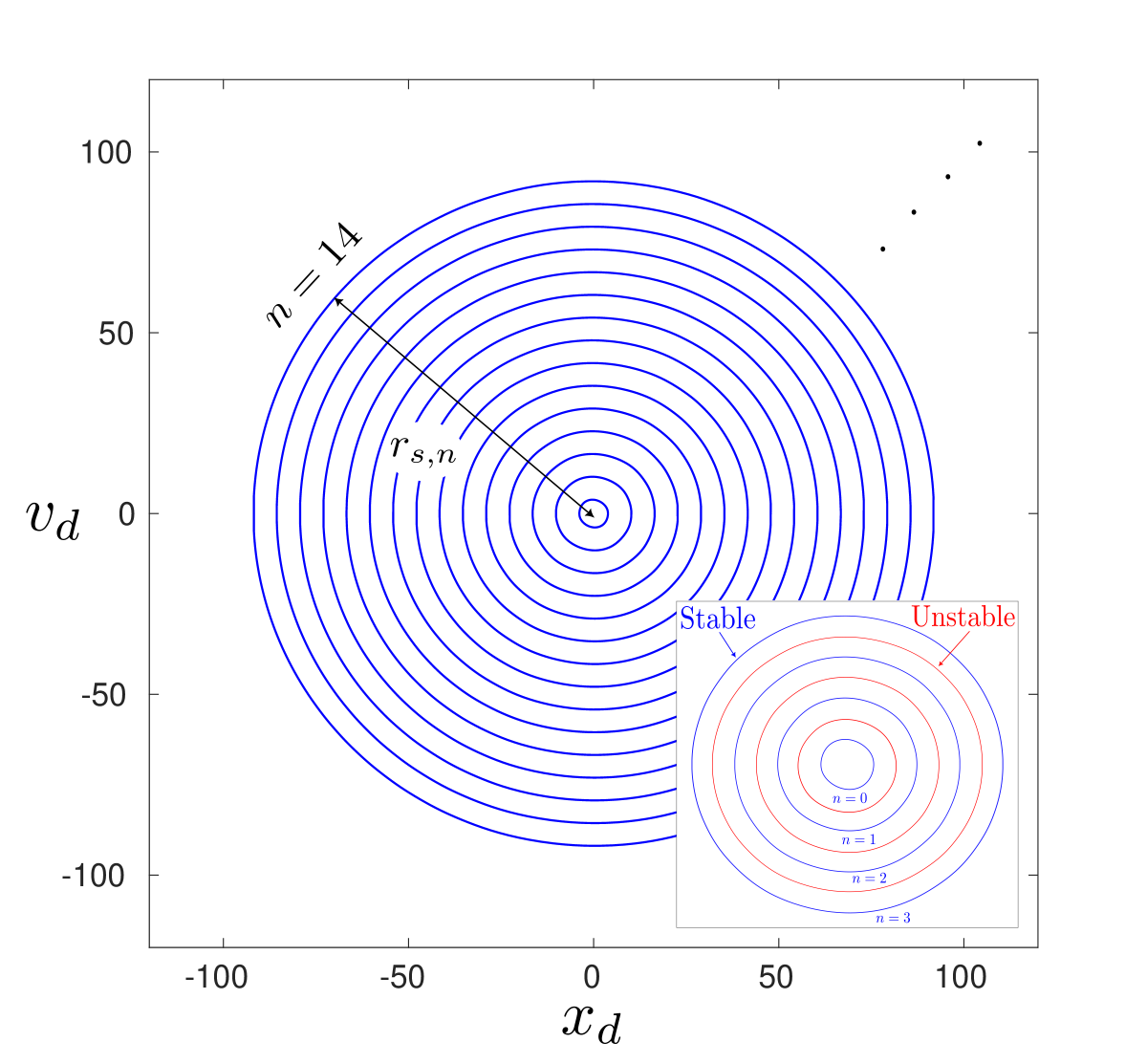}
\caption{Orbit spectrum of the WPE in the low-memory self-oscillator model (dots indicate that the sequence continues). A series of fifteen concentric limit cycles for $\mu=0$ and $\epsilon=0.5$ displaying a countably infinite megastable set of quantized orbits in the phase space formed by the dimensionless position $x_d$ and dimensionless velocity $v_d$ of the particle, shown in Eq.~\eqref{eq:5}. The inset shows the alternating structure of stable (blue) and unstable (red) limit cycles with the first four shown. The unstable limit cycles also form smooth basin boundaries separating the basin of attraction of stable limit cycles.}
\label{Fig:1}
\end{figure}
 
 In the more general case $\bar{\mu}>0$, there is only a finite number $N$ of limit cycles. Solving for $\dot{r}=0$, and using again the asymptotic expansion of the Bessel function, we obtain the transcendental equation $r^{3/2}=\epsilon /\bar{\mu}\sqrt{\pi/2} \cos(r-3 \pi/4)$. For $r \gg 1$, the maximum number of limit cycles is restricted to $N=\lfloor \delta/2 \pi-3/8\rfloor$, where $\delta=(\pi/2)^{1/3}(\epsilon/\bar{\mu})^{2/3}$ and $\lfloor \cdot \rfloor$ is the floor function. This gives us a scaling relation for the number of stable limit cycles in the form $N(\epsilon,\mu) \propto (\epsilon/\mu)^{2/3}$. Among this spectrum of limit cycle states, the fundamental (first) level corresponds to a self-excited attractor since it shares its basin of attraction with the unstable equilibrium point $(x^{*}_d,v^{*}_d)=(0,0)$, while the remaining higher quantized levels are hidden attractors~\cite{KUZNETSOV20145445}, since they do not share their basin of attraction with any equilibrium point. As the dissipation coefficient $\mu$ increases, an infinite cascade of subcritical saddle-node bifurcations take place disintegrating the megastable structure of periodic orbits in decreasing order of their size.
 
 In Fig.~\ref{Fig:1} we have numerically computed the limit cycles of Eq.~\eqref{eq:5} beyond the first ten levels, using constant initial conditions $(x_d(0),v_d(0))=(A,0)$, with increasing values of $A$ and fixed values of $\mu=0$ and $\epsilon=0.5$. Numerical simulations confirm the prediction of a megastable structure and for $n \gg 1$, the limit-cycle size grows following an asymptotically linear trend with $n$. 
 

To characterize the nature of the energy spectrum, we compute the quantization relation for both the energy and the frequency of quantized states. Motivated by recent connections between self-excited oscillators and the quantum potential~\cite{onanelec,bohm52}, we define the energy of our oscillator using the Lyapunov function associated with Eq.~\eqref{Eq: dimless eq}, when $\epsilon=0$. This function describes the conservative energy content of the oscillator~\citep{LOPEZ2023113412,Erneu23}, and can be written as
\begin{equation}
E(x_d,v_d) = \frac{1}{2} v^2_d + \frac{1}{2} x^2_d.
\label{eq:8}
\end{equation}
As opposed to Hamiltonian dynamical systems, for which energy $E$ is a constant of motion and corresponds to fine-tuned parameters $\mu=0$ and $\epsilon=0$, self-excited oscillators gain and lose mechanical energy during different parts of the oscillation, forming a thermodynamic engine~\cite{lopezte}. The non-conservative self-force on the right hand side of Eq.~\eqref{Eq: dimless eq} implies that energy is gained by the oscillator during some part of the limit cycle, while energy is released during the complementary part, even when linear dissipation is absent (\emph{i.e.} $\mu=0$). The addition of linear dissipation disrupts this balance and ceases the existence of larger orbits. These non-conservative forces are related to the time-reversal asymmetry of Eqs.~\eqref{Eq: dimless eq} and~\eqref{eq:5}, and, more generally, of self-excited systems \cite{jenkins,mackey2011}. The resulting dissipative structure~\cite{Pri78} constitutes an active, open dynamical system, which \emph{conserves its mechanical energy on average}. This can be shown by computing the Melnikov function~\cite{Davidow2017} for Eq.~\eqref{eq:5} along the megastable quantized orbits, which corresponds to an averaging of Eq.~\eqref{eq:62} over $\theta$, 
and showing that it takes a zero value~(see Appendix~G). 
\begin{figure*}
\centering
\includegraphics[width=1.8\columnwidth]{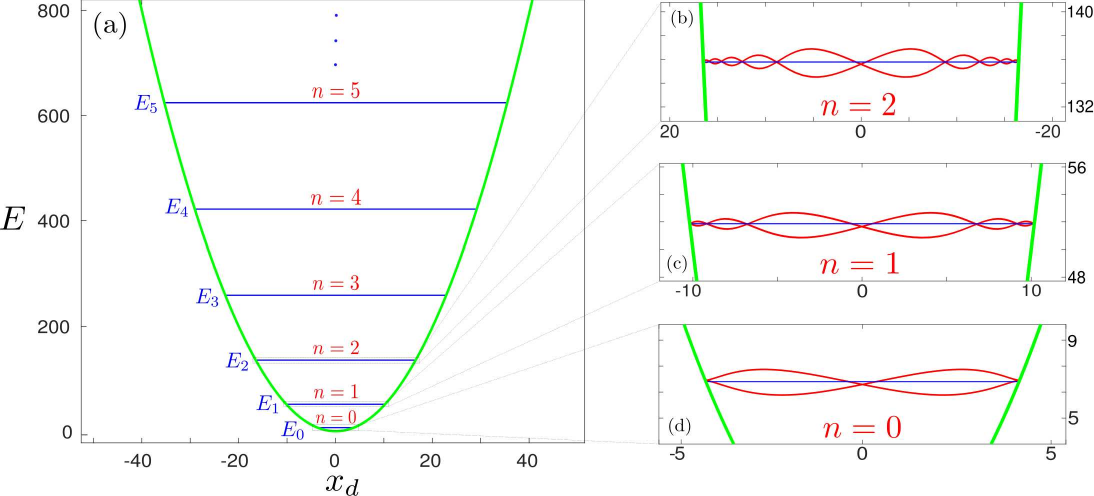}
\caption{Energy spectrum of the WPE in the low-memory self-oscillator model (dots indicate that the sequence continues). (a) The first six mean Lyapunov energy levels $E_n$ (blue) in the external harmonic well (green), together with the time evolution of the Lyapunov energy function $E(x_d,v_d)$ (b-d) for orbits in the zero-point energy level ($n=0$) and the two first excited levels ($n=1$ and $n=2$). The energy oscillations with dimensionless particle position $x_d$ (see Eq.~\eqref{eq:5}) along the quantized orbits (red) appear to obey a mathematical relation for the number of intersections i.e. nodes, with $4n+3$ nodes for $n$th energy level.}
\label{Fig:2}
\end{figure*}

\begin{figure}
\centering
\includegraphics[width=0.9\columnwidth]{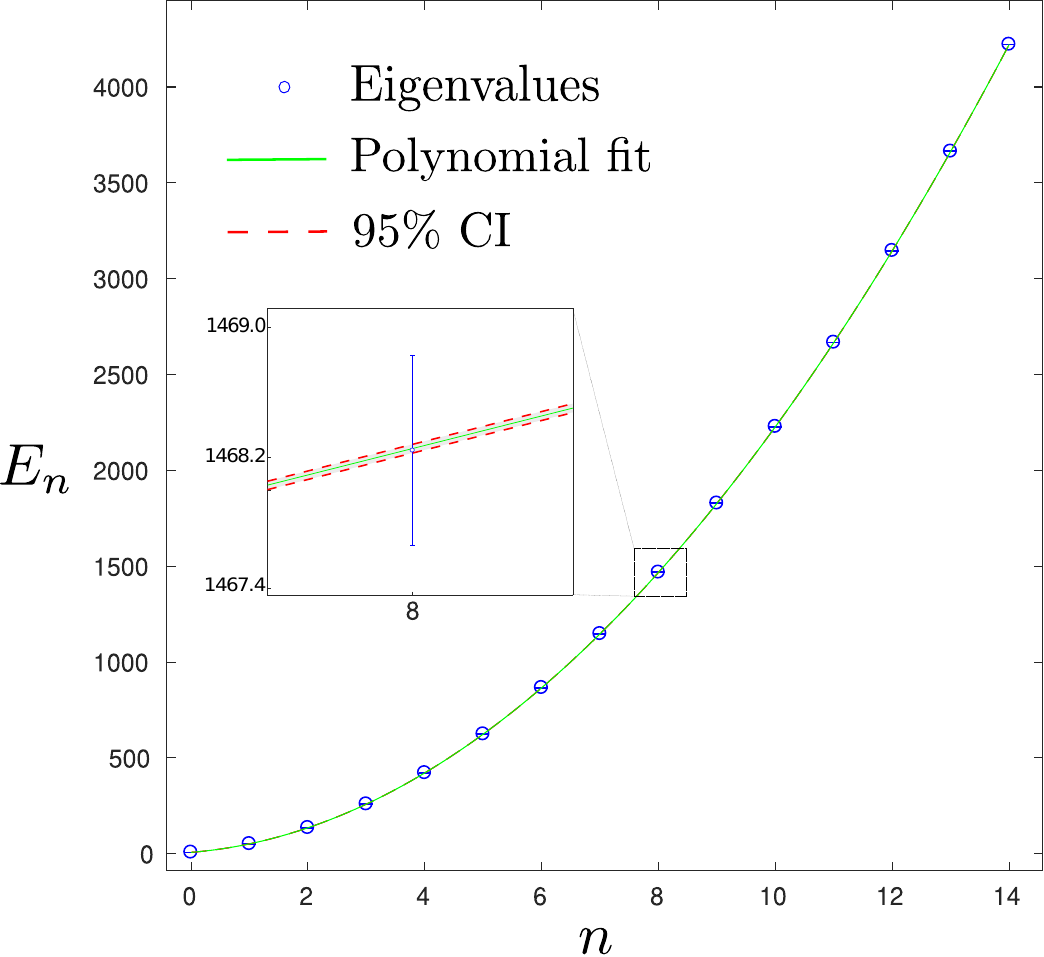}
\caption{The energy levels of the discrete spectrum follow a quadratic trend. The least-squares fitting is very accurate, with tight 95 \% confidence intervals. The error bars (blue) represent the variation in energy uncertainty along the quantized orbit, computed as the variance of the Lyapunov energy function.
}
\label{Fig:3}
\end{figure}

Consequently, we assign an energy eigenvalue for each stable limit cycle, 
 $E_n=(1/T)\int^T_0 E(x_{d,n}(t),v_{d,n}(t))dt$, where $(x_{d,n}(t),v_{d,n}(t))$ represent the trajectory of the $n$th stable limit cycle with period $T$. These energy levels within the external harmonic potential are depicted in Fig.~\ref{Fig:2}. Since the stable limit cycles are approximately circular, with radius size $r_{s,n} \approx \pi (2 n + 5/4)$, we can estimate the energy spectrum of Eq.~\eqref{eq:5} as $E_n \approx r^2_{s,n}/2 = 2 \pi^2 (n + 5/8)^2 $, obtaining a quadratic relation for the energy eigenvalues as a function of the orbit number $n$. This is shown in Fig.~\ref{Fig:3}, where a least squares fitting to the quantization rule $E_n=a n^2+b n+c$ gives the parameter values $a=19.74$, $b=24.70$ and $c=7.42$. These values are in excellent agreement with the theoretical approximations of $a=2\pi^2$, $b=5\pi^2/2$ and $c=25\pi^2/32$, with $c$ corresponding to the zero-point energy of the oscillator.
 The excellent agreement of analytical approximation is a consequence of the quasi-harmonic approximation of the limit cycles, due to the smallness of the parameter $\epsilon$.  

The period or frequency of the limit cycles remains almost constant for the entire megastable set and does not vary significantly with the orbit number $n$. 
The Fourier analysis of the periodicity of the spectrum of the limit cycles set is quasi-monochromatic with an average asymptotic value of $\omega_n\approx0.995$, nearly coinciding with the predicted value of $\omega_n=1$ for the unperturbed harmonic oscillator. This constitutes a fundamental difference with respect to a quantum harmonic oscillator, for which both the energy and the frequency spectra are proportional to each other. Nevertheless, we suggest the possibility that different energy-frequency quantum-like functional relations may arise in our system for other nonlinear confining potentials $V(x)$ and/or different wave forms $W(x)$ of the WPE. 


\textit{Discussion and conclusions.} Experimental and theoretical works with the walking-droplet system have shown the existence of a finite spectrum of quantized states with at most a few coexisting limit cycle states~\citep{Perrard2014a,Perrard2014b,Tambasco2016,Tambascoorbit,Labousse_2014,labousseharmonic,Kurianskiharmonic,durey2018,Perrard2018}. Our results indicate that an extended set of quantized orbits might be achievable in a confining one-dimensional potential for very small dissipation. In addition to the simplified model presented here, we obtain megastability in both the continuous stroboscopic model with truncated memory of duration equal to one bouncing period i.e. $\tau=1$~(see Appendixes~B and C), as well as the discrete model~(see Appendixes~A and C) of \citet{Oza2013} with one bounce truncated. In the standard stroboscopic model of \citet{Oza2013} with $\tau=0$, we don't obtain megastable orbits and the particle continuously climbs up the harmonic potential. Although the relevance of our truncated-memory pilot-wave model to the experimental walking-droplet system is presently unclear, nevertheless, it would be interesting to perform experiments with a walking droplet in a one-dimensional harmonic potential in the very low-dissipation regime to either confirm or rule out megastability. It might be possible to generate the $1$D harmonic potential in experiments by injecting the droplet with ferrofluid and applying an external magnetic field as demonstrated by \citet{Perrard2014a}, whereas the low-dissipation regime might be difficult to achieve. 
For typical experimental set-up with silicone oil droplets, we get the value of dimensionless wave force coefficient $\epsilon \sim 0.01 - 0.1$ and the dimensionless drag coefficient $\mu \sim 0.1$. One would need to reduce this drag coefficient by two or three orders of magnitude to obtain an appreciable number of megastable orbits. 
These limitations might perhaps be circumvented using alternate fluids, reducing the parameter $\mu$ to almost zero. For example, the Faraday instability has been observed in superfluids~\citep{PhysRevLett.98.095301}, suggesting the possibility of constructing a walking droplet setup for superfluids where the low-dissipation regime might be realized. Independent of experiments, our work contributes to the framework of generalized pilot-wave dynamics~\citep{Bush2015}, which has been formulated to encourage the exploration of hydrodynamic quantum analogs in a wider class of dynamical systems and parameter-space that are not restricted by experimental constraints. In this context, we show that a walking-droplet inspired pilot-wave model with truncated memory can exhibit megastable quantization in a confining potential.



 We suspect that the megastable structure observed in our system
 is not unique, and it may be possible to obtain megastability more ubiquitously for a harmonic oscillator perturbed by nonlinear and/or time-delayed non-conservative forces. For example, we have also demonstrated the existence of a megastable structure of orbits in a time-delayed system describing a particle in a harmonic potential with a state-dependent time-delayed force, inspired from a retarded self-oscillator recently found in the context of classical electrodynamics of extended bodies~\cite{LOPEZ2023113412} (see Appendix H). Further, the formalism presented here can be extended more generally to perturbed harmonic oscillators of the form
\begin{equation}
\ddot{x}- \epsilon f(\dot{x}) + x= 0.    
\end{equation}
Any function $f(\dot{x})$ that can be expressed as the product of an even periodic function and an odd polynomial (or vice-versa), can provide megastable dissipative structures (see Appendix F). The Krylov-Bogoliubov method can be applied in general to such near-Hamiltonian systems~\citep{doi:10.1142/S0218127407018208} 
allowing one to compute the maximum number of stable limit cycles and determine the existence of megastable structures~\cite{giac97}.


\textit{Acknowledgments}. R.V. acknowledges the support of the Leverhulme Trust [Grant No. LIP-2020-014]. 



\bibliography{megastable_droplets}

\appendix

\section*{Appendixes}

\section{Derivation of low-memory self-excited oscillator model}
We start with the following one-dimensional truncated-memory stroboscopic model for the wave-particle entity (WPE): 
\begin{align}\label{Eq: dimensional eq supp}
    &m\ddot{x}_d+D\dot{x}_d + k x_d = \\ \nonumber
    &-\frac{F}{T_F}\int_{-\infty}^{t-t_c} W'\left(k_F(x_d(t)-x_d(s))\right)\,\text{e}^{-\frac{(t-s)}{T_F \text{Me}}}\,\text{d}s. 
\end{align}
Here $m$ is the droplet mass, $D$ is an effective time-averaged drag coefficient, $k$ is the spring constant of the external harmonic potential, $T_F$ is the period of Faraday waves and also the bouncing period of the droplet, $\text{Me}$ is the memory parameter and $F=mg A k_F$ where $A$ is the amplitude of surface waves, $g$ is gravitational acceleration and $k_F$ is the wave number associated with droplet-generated damped Faraday waves~(see \citep{Oza2013} for more details on these parameters). The physical relevance of the cut-off time $t_c$ lies in the fact that it takes the droplet one more bounce to feel the effect of its most recent impact, since the droplet is in-flight right after the underlying field is reshaped by such an impact. This model reduces to the one-dimensional standard stroboscopic model of \citet{Oza2013} when the memory cut-off time $t_c=0$. The standard continuous stroboscopic model was derived as an approximation of the discrete impact model given by
\begin{align}\label{Eq: dimensional eq discrte old}
    &m\ddot{x}_d+D\dot{x}_d + k x_d = \\ \nonumber
    &-{F}\sum_{n=-\infty}^{\lfloor t/T_F \rfloor} W'\left(k_F(x_d(t)-x_d(n T_F))\right)\,\text{e}^{-\frac{(t-n T_F)}{T_F \text{Me}}}. 
\end{align}
In this model, the droplet generates waves at discrete times $t=n T_F$ and the wave memory force on the droplet is calculated by taking into account all the wave including the most recent bounce at time $\lfloor t/T_F \rfloor$ where $\lfloor \cdot \rfloor$ is the floor function. Although we have not found megastability in both the continuous stroboscopic model with $t_c=0$ in Eq.~\eqref{Eq: dimensional eq supp} and the discrete model in Eq.~\eqref{Eq: dimensional eq discrte old}, we observe that megastability is robust~(see Sec.~\ref{comparison of models} of this Appendix) in both the truncated continuous stroboscopic model with $t_c=T_F$ as well as the discrete stroboscopic model in Eq.~\eqref{Eq: dimensional eq discrte} when the summation is truncated one bounce earlier i.e. at $\lfloor t/T_F \rfloor -1$ giving
\begin{align}\label{Eq: dimensional eq discrte}
    &m\ddot{x}_d+D\dot{x}_d + k x_d = \\ \nonumber
    &-{F}\sum_{n=-\infty}^{\lfloor t/T_F \rfloor -1} W'\left(k_F(x_d(t)-x_d(n T_F))\right)\,\text{e}^{-\frac{(t-n T_F)}{T_F \text{Me}}}. 
\end{align}
We non-dimensionalize both the truncated-memory continuous and discrete model by using $x'=k_F x$ and $t'=t/T_F$ and drop primes on the dimensionless variables giving
\begin{align}\label{Eq: dimeless cont supp}
    &\ddot{x}_d+\mu\dot{x}_d + K x_d = 
    -\beta\int_{-\infty}^{t-\tau} W'\left(x_d(t)-x_d(s)\right)\,\text{e}^{-\frac{(t-s)}{ \text{Me}}}\,\text{d}s. 
\end{align}
and
\begin{align}\label{Eq: dimless disc supp}
    &\ddot{x}_d+\mu\dot{x}_d + K x_d = \\ \nonumber
    &-\beta\sum_{n=-\infty}^{\lfloor t-1 \rfloor} W'\left(k_F(x_d(t)-x_d(n ))\right)\,\text{e}^{-\frac{(t-n)}{\text{Me}}}. 
\end{align}
Here, $\mu=DT_F/m$, $K=k T_F^2/m$, $\tau=t_c/T_F$ and $\beta=F k_F T_F^2/m$ are the dimensionless drag coefficient, spring constant, cut-off memory time and wave force coefficient, respectively. Now, we consider a low-memory limit where the particle motion is only influenced from its most recently generated wave at time $t-\tau$ with $\tau=1$. Thus, keeping only the contribution from the integral at $t-\tau$, we get the delayed differential equation
\begin{align}\label{eq: single delay}
\ddot{x}_d+\mu\dot{x}_d+K x_d=-\beta\,W'({x}_d(t)-x_d(t-1))\,\text{e}^{-1/\text{Me}}.
\end{align}
Further approximating by assuming that the speed is constant in this short time interval of $\tau=1$, we have $x_d(t)-x_d(t-1)\approx \dot{x}_d(t)$, and we get 
\begin{align}
\ddot{x}_d+\mu\dot{x}_d+K x_d=-\epsilon W'(\dot{x}_d),
\label{eq:sm_1}
\end{align}
where $\epsilon=\beta\,\text{e}^{-\frac{1}{\text{Me}}}$. This results in our low-memory self-excited oscillator equation. We note that the above equation with $K=0$ is same in structure to the seminal model of a walking droplet at low-memory proposed by \citet{Protiere2006}.

\begin{figure*}
\centering
\includegraphics[width=2.0\columnwidth]{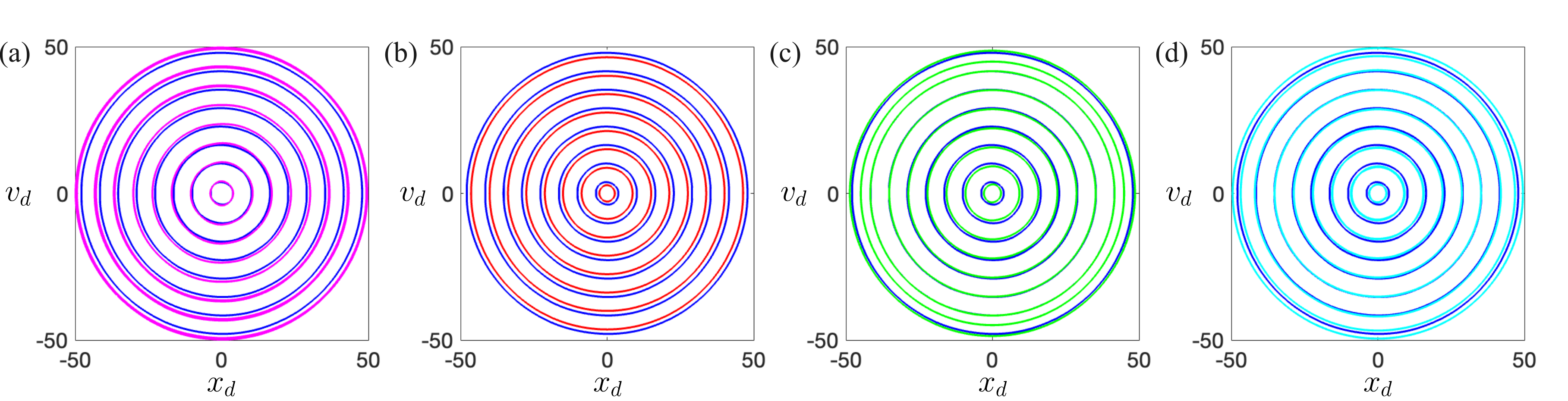}
\caption{Comparison of megastable set of orbits for sinusoidal wave form $W(x)=\cos(x)$ between the low-memory self-oscillator model of Eq.~\eqref{eq:sm_1} presented the main text (blue) and (a) low-memory single delay model in Eq.\eqref{eq: single delay} (magenta), (b) Lorenz-like ODE model in Eq.\eqref{Lorenz ODE trucated} (red), (c) Lorenz-like delay model in Eq.\eqref{Lorenz delay truncated} (green) and (d) discrete truncated-memory model in Eq.\eqref{Eq: dimless disc supp} (cyan). Parameter values were fixed to $\tau=1$, $K=1$, $\mu=0$, $\text{Me}=1$ and $\epsilon=\beta\,\text{e}^{-1/\text{Me}}=0.5$. The first eight orbits of the megastable spectrum are shown, for clarity.}
\label{Fig: Megastability compare}
\end{figure*}

\section{Lorenz-like systems from truncated-memory stroboscopic model}

It has been shown that the continuous stroboscopic model of \citet{Oza2013} for a one-dimensional WPE with a sinusoidal wave form i.e. $W(x)=\cos(x)$, reduces to the celebrated Lorenz equations~\citep{Durey2020lorenz,ValaniUnsteady,Valanilorenz2022}. Here, we show that the truncated stroboscopic model of Eq.~\eqref{Eq: dimeless cont supp} for a sinusoidal wave form can also be transformed to a Lorenz-like dynamical system. For a cosine waveform, $W(x)=\cos(x)$, Eq.~\eqref{Eq: dimeless cont supp} transforms to
\begin{align}\label{Eq: dimeless cont supp sine}
    &\ddot{x}_d+\mu\dot{x}_d + K x_d = 
    \beta\int_{-\infty}^{t-\tau} \sin\left(x_d(t)-x_d(s)\right)\,\text{e}^{-\frac{(t-s)}{ \text{Me}}}\,\text{d}s. 
\end{align}
Let $$Y(t)=\beta\int_{-\infty}^{t-\tau} \sin\left(x_d(t)-x_d(s)\right)\,\text{e}^{-\frac{(t-s)}{ \text{Me}}}\,\text{d}s,$$ then, using Leibniz rule for differentiation under the integral sign, we get
\begin{align*}
  \dot{Y} = -\frac{1}{\text{Me}} Y + v_dZ + \beta\,\text{e}^{-\tau/\text{Me}}\,\sin(x_d(t)-x_d(t-\tau))  
\end{align*}
where $\dot{x}_d=v_d$ and $$Z=\beta\int_{-\infty}^{t-\tau} \cos\left(x_d(t)-x_d(s)\right)\,\text{e}^{-\frac{(t-s)}{ \text{Me}}}\,\text{d}s.$$ Differentiating $Z$ with respect to time in a similar way we get
\begin{align*}
  \dot{Z} = -\frac{1}{\text{Me}} Z - v_d Y + \beta\,\text{e}^{-\tau/\text{Me}}\,\cos(x_d(t)-x_d(t-\tau)). 
\end{align*}
Hence, the stroboscopic model in Eq.\eqref{Eq: dimeless cont supp} reduces to the following Lorenz-like system with time delay
\begin{align}\label{Lorenz delay truncated}
    \dot{x}_d&=v_d,\\ \nonumber
    \dot{X}&= Y -\mu v_d - K x_d,\\ \nonumber
    \dot{Y}&=-\frac{1}{\text{Me}} Y + v_d Z + \beta\,\text{e}^{-\tau/\text{Me}}\,\sin(x_d(t)-x_d(t-\tau)),\\ \nonumber
    \dot{Z}&= -\frac{1}{\text{Me}} Z - v_d Y + \beta\,\text{e}^{-\tau/\text{Me}}\,\cos(x_d(t)-x_d(t-\tau)).
\end{align}
The above set of equations can be further reduced to a set of ordinary differential equations (ODEs) by approximating $x_d(t)-x_d(t-\tau)\approx \tau v_d(t)$ for small $\tau$ by assuming that the velocity of the particle stays almost constant between $t-\tau$ and $t$, yielding
\begin{align}\label{Lorenz ODE trucated}
    \dot{x}_d&=v_d,\\ \nonumber
    \dot{X}&= Y -\mu v_d - K x_d,\\ \nonumber
    \dot{Y}&=-\frac{1}{\text{Me}} Y + v_d Z + \beta\,\text{e}^{-\tau/\text{Me}}\,\sin(\tau v_d),\\ \nonumber
    \dot{Z}&= -\frac{1}{\text{Me}} Z - v_d Y + \beta\,\text{e}^{-\tau/\text{Me}}\,\cos(\tau v_d).
\end{align}


\section{Comparison of megastability in different models}\label{comparison of models}

Although we used the analytically tractable low-memory self-excited oscillator model in the main text to demonstrate megastability, we note that the phenomenon of megastability is robust in the truncated-memory continuous and discrete models. For this purpose, we compare the megastable orbits for the wave form $W(x)=\cos(x)$ between the low-memory self-oscillator ODE model and (i) low-memory single delay model in Eq.~\eqref{eq: single delay}~(see also Supplemental Videos S1-S3), (ii) the continuous truncated-memory stroboscopic model that reduces to a Lorenz-like ODE system in Eq.~\eqref{Lorenz ODE trucated}, (iii) the continuous truncated-memory stroboscopic model that reduces to a Lorenz-like system with time-delay as in Eq.~\eqref{Lorenz delay truncated}~(see also Supplemental Videos S4-S6) and (iv) the discrete truncated-memory model in Eq.~\eqref{Eq: dimless disc supp}. A comparison of the first $8$ megastable orbits between these models is shown in Fig.~\ref{Fig: Megastability compare}. As we can see, all the models exhibit the phenomenon of megastability, although the quantitative details of the positioning of the levels and the spacing between them vary between the models.

\begin{figure*}
\centering
\includegraphics[width=2.0\columnwidth]{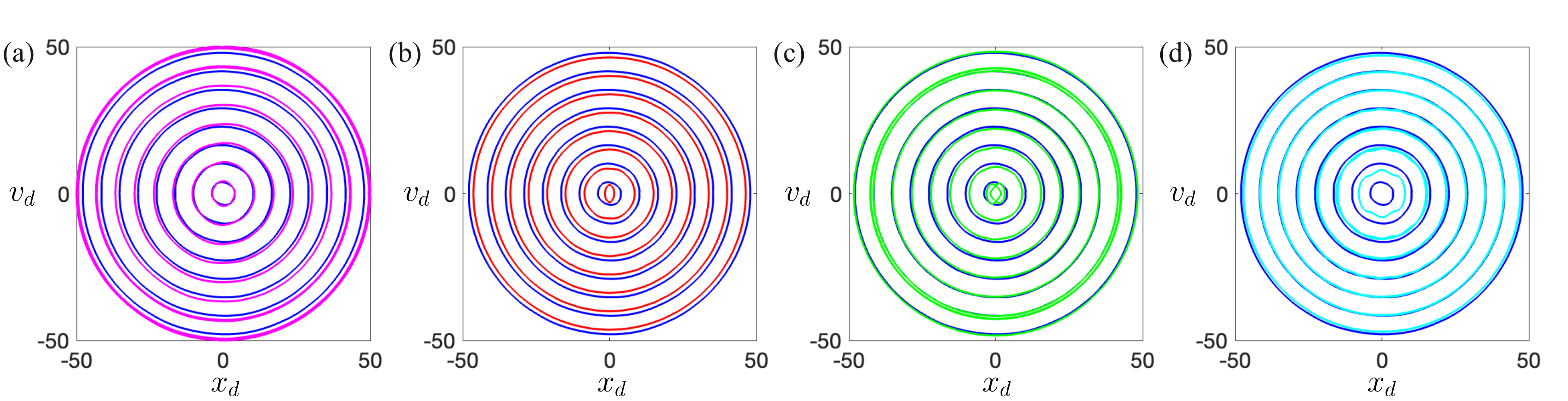}
\caption{Comparison of megastable set of orbits for a sinusoidal wave form $W(x)=\cos(x)$ between the low-memory self-oscillator model of Eq.~\eqref{eq:sm_1} presented the main text (blue) and (a) low-memory single delay model in Eq.\eqref{eq: single delay} (magenta), (b) the Lorenz-like ODE model in Eq.\eqref{Lorenz ODE trucated} (red), (c) the Lorenz-like delayed model appearing in Eq.\eqref{Lorenz delay truncated} (green) and (d) discrete truncated-memory model represented by Eq.\eqref{Eq: dimless disc supp} (cyan). The parameter values have been fixed to $\tau=1$, $K=1$, $\mu=0$, $\text{Me}=10$ and $\beta=1$.}
\label{Fig: Megastability compare high memory}
\end{figure*}

Furthermore, to show that megastability is not strictly limited to the low-memory regime, we plot the megastable set of orbits of the four different models for a larger value of the memory $\text{Me}=10$ and $\beta=1$ in Fig.~\ref{Fig: Megastability compare high memory}. As it is clear from the figures, although the geometry of the orbits gets distorted, the discrete structure of megastable the spectrum is robust. 


\section{Averaged equation of the amplitude for a Bessel wave field}

The Bessel wave field with $W(x)=J_0(x)$, where $J_0(x)$ is the Bessel function of zero order and first kind, is routinely used in pilot-wave models of a walking droplet~\citep{Oza2013} since it more accurately captures the experimental waves generated by the droplet (due to the presence of spatial decay in the Bessel function). If we use this wave form in our self-oscillator model then we get
\begin{align*}
\ddot{x}_d+\mu\dot{x}_d+K x_d=\epsilon J_1(\dot{x}_d),
\end{align*}
where $J_1(x)$ is the Bessel function of the first order and first kind. In this case, we can use the Bessel function Taylor series expansion 
\begin{align*}
J_{1}(x)= \sum^{\infty}_{k=0} \frac{(-1)^n}{k!(k+1)!} \left( \frac{x}{2} \right)^{2k+1},
\end{align*}
and apply the Krylov-Bogoliubov method~\citep{krylov1950} to obtain the following averaged equation for the radius of the limit cycles:
\begin{align*}
\dot{r} & = - \frac{\mu}{2 \pi} r \int^{2\pi}_0 \cos^2\theta d\theta \\ \nonumber 
& + \frac{\epsilon }{2 \pi} \sum^{\infty}_{k=0} \frac{(-1)^k}{k!(k+1)!} \left(\frac{r}{2}\right)^{2 k+1} \int^{2\pi}_0 \cos^{2 k + 2}\theta d\theta. 
\end{align*}
The integral appearing in the above equation can be computed in terms of Euler's Gamma function $\Gamma(x)$. The result is
\begin{align*}
\dot{r} & = - \frac{\mu}{2} r \\ \nonumber
& + \frac{\epsilon }{2 \pi} \sum^{\infty}_{k=0} \frac{(-1)^k}{k!(k+1)!}  \left(\frac{r}{2}\right)^{2 k+1} \frac{2 \sqrt{\pi} \Gamma(k+3/2)}{\Gamma(k+2)}. 
\end{align*}
Recalling the relations $\Gamma(n+1)=n!$ and $\Gamma(n+1/2)=(2 n)! \sqrt{\pi}/(2^{2n} n!)$, yields the series
\begin{align*}
\dot{r} & = - \frac{\mu}{2} r \\ \nonumber
& + \frac{\epsilon}{2} \sum^{\infty}_{k=0} (-1)^k \left(\frac{r}{4}\right)^{2 k+1} \frac{ (2k+2)!}{k! [(k+1)!]^3}. 
\end{align*}
The above series converges to a product of Bessel functions, and can be expressed as
\begin{align}
\dot{r} & = - \mu r + \epsilon J_{0}(r) J_{1}(r), \label{eq:41x4} 
\end{align}
where we have rescaled $(\mu,r) \rightarrow (2 \mu, 2 r)$ to simplify the equation further.

Thus, for $\mu=0$, we obtain a megastable structure with infinite limit cycles given by the zeros of the product of Bessel functions.

\section{Averaged equation of the amplitude for a sinusoidal wave field with spatial decay}

Although the Bessel function is a more realistic wave form for experiments compared to a pure sinusoidal wave form, it still does not accurately capture the exponential spatial decay observed in experiments~\cite{Damiano2016}. To capture this whilst maintaining a smooth and differentiable wave field, we consider the wave form, $W(x)=\cos(x)\,\text{sech}(x/2l)$, which results in the following self-oscillator model
\begin{align}
 \label{eq:411}
&\ddot{x}_d+\mu\dot{x}_d+K x_d=\\ \nonumber
&\epsilon \sin(\dot{x}_d) \text{sech}\left(\frac{\dot{x}_d}{2l}\right)+\frac{\epsilon}{2l}\cos(\dot{x}_d) \text{sech}\left(\frac{\dot{x}_d}{2l}\right)\text{tanh}\left(\frac{\dot{x}_d}{2l}\right), 
\end{align}
where $l$ is a dimensionless length of spatial decay (scaled by the Faraday wavenumber $k_F$).

When the spatial decay is sufficiently slow i.e. $l\gg1$, we can use asymptotics to do a Taylor series expansion of the hyperbolic functions to second order, and obtain
\begin{align}
 \label{eq:51}
&\ddot{x}_d+\mu\dot{x}_d+K x_d=\\ \nonumber
&\epsilon \sin(\dot{x}_d) +\frac{\epsilon}{4l^2}\cos(\dot{x}_d)\dot{x}_d-\frac{\epsilon}{8l^2}\sin(\dot{x}_d)\dot{x}^2_d. 
\end{align}
Utilizing the Krylov-Bogoliubov method, the three self-excited terms appearing on the right hand side of Eq.~\eqref{eq:51} can be expressed as a superposition of Bessel functions in the form
\begin{align}
 \label{eq:51x2}
\dot{r}=&-\mu r + \frac{\epsilon}{8 l^2} ((4+8 l^2) J_{1}(r)- 7 r J_{2}(r) \\  \nonumber
&+ r^2 J_3(r)),      
\end{align}
where, again, we have rescaled the drag parameter $\mu \rightarrow 2 \mu$, for simplicity. We note that in the limit $l \rightarrow \infty$ we recover the cosine profile, as expected.

\begin{figure}
\centering
\includegraphics[width=1.0\columnwidth]{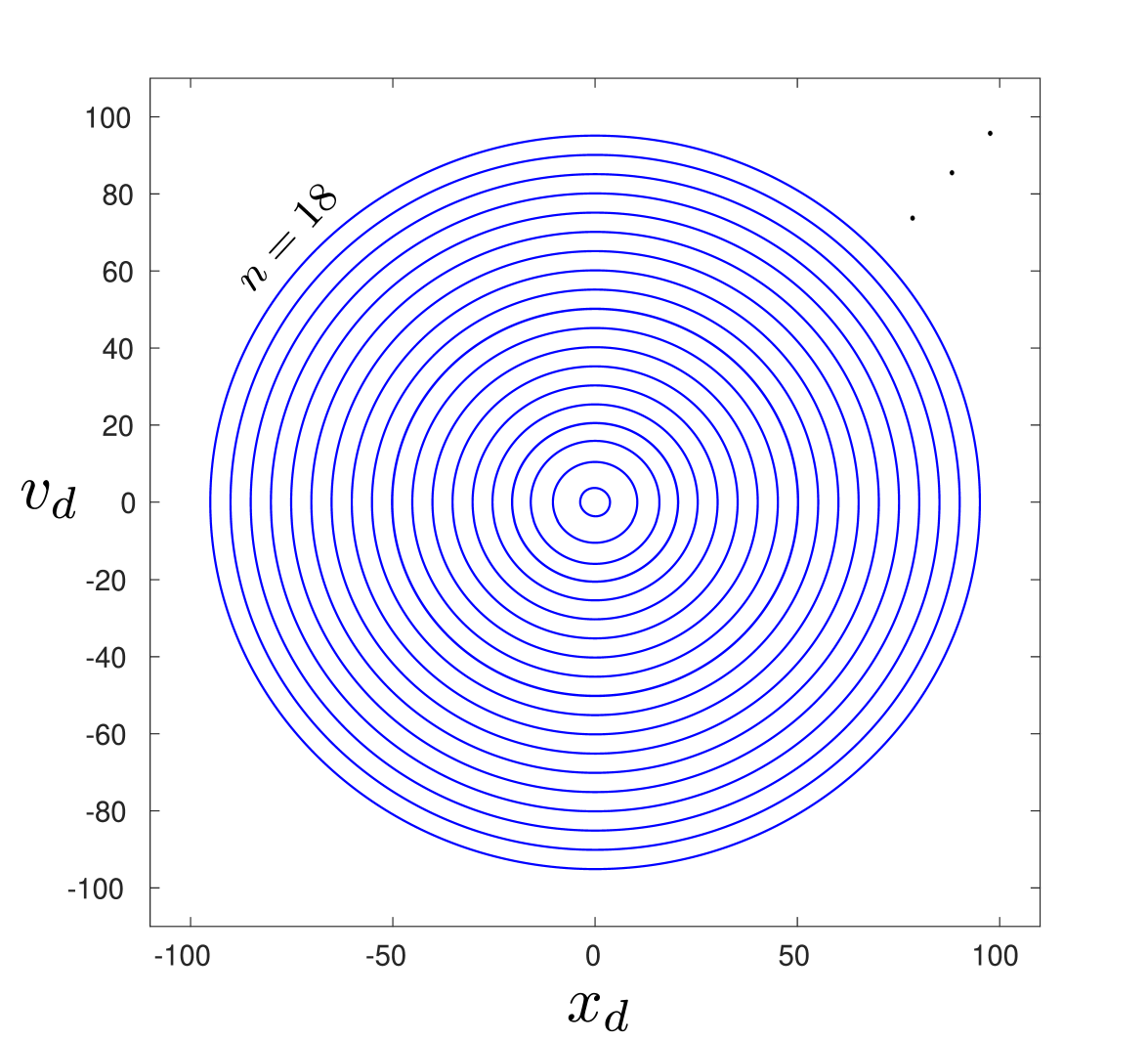}
\caption{Megastable structure of orbits for Eq.~\eqref{eq:411} with parameter values $K=1$, $\mu=0$, $\epsilon=0.5$ and $l=1$, which corresponds to a typical experimental value for the spatial decay rate of the droplet generated pilot-wave.}
\label{Fig:1s}
\end{figure}

Thus, we again obtain a megastable structure given by the zeros of the above combination of Bessel functions. Further, it is interesting to point out that even in the limit of small $l$, we retain the megastable structure from Eq.~\eqref{eq:411}. An example is shown in Fig.~\ref{Fig:1s} for $l=1$ which would correspond to the decay length of a typical walking-droplet experimental setup. As far as we have computed, these megastable structures exist for values as small as $l=0.01$, even though the first energy level becomes increasingly distorted. This leads us to conjecture that a monotonically decreasing spatial profile multiplied by the oscillatory wave form does not destroy the megastable set of orbits. Moreover, we also find that for $\mu=0$ and larger memory, megastability still exists in the truncated-memory stroboscopic model of Eq.~\eqref{Eq: dimeless cont supp} for a sinusoidal wave form with spatial decay. 

\section{A theorem for the existence of megastability in general self-excited oscillators}

We now consider a more general nonlinear self-excited oscillator
\begin{align*}
\ddot{x}_d-\epsilon f(\dot{x}_d)+x_d=0,
\end{align*}
where the function $f$ is the product of an odd polynomial multiplied by an even trigonometric function. The proof of the reciprocal case of the product of an even polynomial and an odd trigonometric function, follows along the same line of reasoning. In the case that the polynomial and the periodic function are both odd or both even, they do not contribute to the averaged equations of the amplitude, but only to the average equation of the phase. Thus, in general $f$ can be the product of any polynomial of arbitrary degree and coefficients, and any periodic function. We start with the following ansatz
\begin{align*}
f(\dot{x})= \dot{x}^{2 s+1} \sum^{\infty}_{k=0} a_k \cos(k \dot{x}),
\end{align*}
and use the Krylov-Bogoliubov method to get the averaged equation of the amplitude $r(t)$ as
\begin{align*}
\dot{r} & = \frac{\epsilon}{2 \pi}  \sum^{\infty}_{k=0} \sum^{\infty}_{n=0}  a_k \frac{(-1)^n k^{2 n}}{(2 n)!}  r^{2 s + 2 n+1} \int^{2\pi}_0 \cos^{2 n + 2 s +2}\theta d\theta. 
\end{align*}
The above integral can be exactly solved in terms of Euler's Gamma function $\Gamma(x)$, yielding the result
\begin{align*}
\dot{r} & = \frac{\epsilon}{2\pi} \sum^{\infty}_{n=0} \sum^{\infty}_{k=0}  a_k \frac{(-1)^n k^{2 n}}{(2 n)!}  r^{2 s + 2 n+1} \frac{2 \sqrt{\pi}\Gamma(n+s+3/2)}{\Gamma(n+s+2)}, 
\end{align*}
which can be expressed in terms of factorials as
\begin{align*}
\dot{r} & =  \frac{\epsilon}{2} \left( \frac{r}{2}\right)^{2 s + 1} \sum^{\infty}_{k=0}  a_k  \sum^{\infty}_{n=0}\frac{(-1)^n}{(2 n)!}  \left( \frac{k r}{2}\right)^{2 n} \frac{(2(n+s)+2)!}{[(n+s+1)!)]^2}. 
\end{align*}
The summation over $n$ can be computed in terms of the regularized generalized hypergeometric functions $_{p}F_{q}(b;c;x)$ as follows
\begin{align}
\label{eq:91} 
& \dot{r} = \epsilon r^{2 s + 1}  \Gamma(s+3/2) \sum^{\infty}_{k=0}  a_k \\ \nonumber
& \left[ _{p}F_{q}(b_1;c_1;-k^2 r^2/4)-k^2r^2/4~_{p}F_{q}(b_2;c_2;-k^2 r^2/4 \right], 
\end{align}
where we have introduced the parameters $b_1=s+1/2$, $c_1=(1/2,s+2)$, $b_2=s+3/2$ and $c_2=(3/2,s+3)$. The convergence of this hypergeometric series is guaranteed by the fact that the coefficients $a_k$ are square-summable, and because the two hypergeometric functions form a complete orthonormal basis. 

Furthermore, since the addition of the two hypergeometric functions appearing in Eq.~\eqref{eq:91} (for $s>0$ and $k>0$) results in a function having an infinite number of zeros, we can generally expect that the averaged equation for the amplitude presents an infinite number of zeros, yielding a megastable structure. This result is verified by numerical simulations.

\section{Average energy conservation along quantized orbits}
Consider a Hamiltonian system with the energy function $H(x_p,v_p )$. The system is opened to the environment and its dynamics is governed by the perturbed Hamilton's canonical equations of motion

\begin{align}
\label{eq:sm_13}
 & \dot{x}_d=\frac{\partial H}{\partial v_d}+g_1(\epsilon,x_d,v_d),\\ \nonumber
 &\dot{v_d}=-\frac{\partial H}{\partial x_d}+g_2(\epsilon,x_d,v_d).
\end{align}

We now show that energy is conserved on average if and only if the Melnikov function is equal to zero. We start by assuming that energy is conserved on the average, what can be mathematically written as
\begin{align}
\label{eq:sm_14}
 \oint d H=0 \implies  \oint \left( \frac{\partial H}{\partial x_d} \dot{x}_d + \frac{\partial H}{\partial v_d} \dot{v}_d \right)d t =0.
\end{align}
Substituting Eqs.~\eqref{eq:sm_13} in Eq.~\eqref{eq:sm_14} we get 
\begin{align}
\label{eq:sm_14}
M(\epsilon) \equiv \oint \left(\frac{\partial H}{\partial x_d}g_1 + \frac{\partial H}{\partial v_d}g_2 \right) d t =0.
\end{align}
Here $M(\epsilon)$ is the Melnikov function. We note that, since quantized orbtis obey Eq.~\eqref{eq:sm_13}, if Eq.~\eqref{eq:sm_14} holds, then $\langle \dot{H} \rangle =0$ is implied. Equivalently, we can express our quantization condition as
\begin{align}
\label{eq:sm_15}
\oint \left(\dot{x}_d g_2  -  \dot{v}_d g_1 \right) d t =0.
\end{align}
Using our dynamical Eq.~\eqref{eq:sm_1}, we have $g_1=0$ and $g_2=-\epsilon W'(\dot{x}_d)$, yielding
\begin{align}
\label{eq:sm_16}
\oint \dot{x}_d W'(\dot{x}_d) d t =0.
\end{align}
The above equation is equivalent to calculating the fixed points of the average equation for the amplitude of the self-oscillator, as done by the Krylov-Bogoliubov method. 

Thus by relating the averaged equation of Krylov-Bogoliubov method and the Melnikov function, we have proved the following theorem: \emph{the quantized orbits correspond to limit cycle trajectories that conserve the mechanical energy on the average.}

\section{Megastability in state-dependent delayed differential equations}

Consider a particle with inertial mass $m$ in a external harmonic potential $V(x)=k x^2/2$ and subjected to a linear drag force with damping coefficient $\zeta$. The particle also has self-interactions and this self-force is represented using a retarded harmonic potential~\cite{lien98,onanelec} $Q(x_{\tau})=\lambda x_{\tau}^2/2$ with state-dependent time-delay, where $x_{\tau}=x(t-\tau(\dot{x}))$.  
The resulting equation of motion is
\begin{equation}
    m\ddot{x}+\zeta \dot{x} + \dfrac{dV}{dx}+\dfrac{dQ}{dx_{\tau}} = 0. 
    \label{eq:1}
\end{equation}
This model is an example of a state-dependent delay system that has been shown to exhibit dynamical analogs of orbit quantization with finite multiple attractors, and analog of quantum tunneling via crisis-induced intermittency~\cite{lopval24}. The amplitude $\lambda$ controls the strength of the delay force on the particle. 

We choose the delay function to take the form $\tau(\dot{x})=\tau_0 \cos^{2}(\sigma \dot{x})$. This functional dependence of time-delay on velocity ensures limited memory and reflection spatial symmetry. The parameter $\tau_0$ represents the maximum value of the time-delay which is attained when the particle velocity is zero. Dividing by $m$ and rescaling $x\rightarrow x/\sigma$, ${\zeta}\rightarrow \zeta/m$, ${k}\rightarrow k/m$ and ${\lambda}\rightarrow\lambda/m$  results in the following equation
\begin{align}
\ddot{x}  + {\zeta} \dot{x} + {k} x + {\lambda} x_{\tau}=0, 
\label{eq:2_3}
\end{align}
with $x_\tau=x(t-\tau(\dot{x}))$ and $\tau(\dot{x})=\tau_0 \cos^2\dot{x}$. 

In the low-memory regime corresponding to small $\tau_0$, a Taylor series expansion of the time-delay term to first order~\cite{Erneu23} transforms Eq.~\eqref{eq:2_3} to the following
\begin{equation}
     \ddot{x}+ (\zeta-\lambda \tau_0 \cos^2\dot{x})\dot{x} + (k+\lambda) x = 0.    
    \label{eq:3}
\end{equation}
This low-memory equation of motion is a nonlinear self-excited oscillator with an oscillatory nonlinear drag term and a rescaled harmonic potential. Introducing $\omega^2=k+\lambda$, $\mu=\zeta-\lambda \tau_0/2$ and $\epsilon=\lambda \tau_0/2$ coefficients results in the following nonlinear oscillator
\begin{equation}
    \ddot{x} + (\mu - \epsilon \cos\dot{x})\dot{x} + x= 0. 
    \label{eq:4}
\end{equation}
where we have rescaled $(x,t,\epsilon,\mu) \rightarrow (x/2\omega, t/\omega, \omega\epsilon, \omega\mu)$ to simplify the equation further. 

In the low-memory limit and the weak dissipation regime ($\zeta \approx \lambda \tau_0/2$), we can assume that $\mu$ and $\epsilon$ are small and apply the Krylov-Bogoliubov averaging method~\cite{krylov1950} to obtain two coupled ordinary differential equations governing the dynamics of the amplitude and the phase of the limit cycles. The resulting averaged equation for the amplitude reads
\begin{equation}
\dot{r}=-\mu r + \epsilon (J_{1}(r)- r J_{2}(r)),      
\end{equation}
where, one last time, we have rescaled parameters $(\mu,\epsilon) \rightarrow (2 \mu, \epsilon/2)$ to fully simplify the equation.

Similar to previous examples, we get infinite zeros of the above equation given by the roots of the combined Bessel functions and we obtain a megastable spectrum.

\section*{Supplementary Videos}

\textbf{Supplemental Video S1-S3:} Video showing system dynamics for $W(x)=\cos(x)$ in the first, second and third megastable orbits, respectively, for low-memory single delay model in Eq.~\eqref{eq: single delay}. In the video, particle position is denoted by a black filled circle, the harmonic potential is shown by a blue curve, the particle-generated wave at time $t-1$ is shown as a red curve (amplitude is exaggerated) and the black curve shows the combined potential that includes both the harmonic potential and the self-generated wave from particle at time $t-1$. The black curve in the videos is mostly overshadowed by the blue curve since the difference between them, which is the particle-generated wave, is very small.
\newline\newline
\textbf{Supplemental Video S4-S6:} Video showing system dynamics for $W(x)=\cos(x)$ in the first, second and third megastable orbits, respectively, for truncated-memory stroboscopic model that takes into account full memory up to time $t-1$ in Eq.~\eqref{Lorenz delay truncated}. In the video, particle position is denoted by a black filled circle, the harmonic potential is shown by a blue curve, the particle-generated wave memory up to time $t-1$ is shown as a red curve (amplitude is exaggerated) and the black curve shows the combined potential that includes both the harmonic potential and the self-generated wave memory. The black curve in the videos is mostly overshadowed by the blue curve since the difference between them, which is the particle-generated wave, is very small.

\end{document}